\newif\ifblind
 \blindfalse

\documentclass{sig-alternate-05-2015}
\usepackage{wasysym}
\usepackage{tikz}
\usepackage{mathptmx}
\usepackage{balance}
\usepackage{booktabs} % table layout
\usepackage{tabulary} % tables with long text
\usepackage{cite}
\usepackage{color}
\usepackage{stackrel}
\usepackage{siunitx}
\usepackage{xspace}   % sophisticated spacing
\usepackage{amsmath}
\usepackage{enumitem} % compact lists
\usepackage{url}
\usepackage{adjustbox}
\newcolumntype{R}[2]{%
    >{\adjustbox{angle=#1,lap=\width-(#2)}\bgroup}%
    l%
    <{\egroup}%
}

\newcommand{\Tdot}{$\CIRCLE$}

\newcommand{\Twdot}{$\Circle$}

\newcommand{\Paragraph}[1]{\smallskip\noindent{\bf #1.}}

\newcommand{\eg}{e.\@\,g.,\@\xspace}

\def\hb{\hbox to 10.7 cm{}}
\def\sharedaffiliation{%
\end{tabular}
\begin{tabular}{c}}

\ifblind
\def\enip{EtherNet/IP\xspace}
\def\Enip{EtherNet/IP\xspace}
\def\swat{BWTT\xspace}
\def\swatlong{Blinded Water Treatment Testbed\xspace}

\def\sutd{UniX\xspace}

\def\SSS{S3\xspace}
\def\SSSlong{SWaT Security Showdown\xspace}
\def\hamids{Detector X\xspace}
\def\hamidsLong{Blinded Monitoring and Intrusion Detection System\xspace}
\def\argus{Detector Y\xspace}

\def\teamone{Team 1\xspace}
\def\teamtwo{Team 2\xspace}
\def\teamthree{Team 3\xspace}
\def\teamfour{Team 4\xspace}
\def\teamfive{Team 5\xspace}
\def\teamsix{Team 6\xspace}
\else
\def\enip{Ethernet/IP\xspace}
\def\Enip{Ethernet/IP\xspace}
\def\swat{SWaT\xspace}
\def\swatlong{Secure Water Treatment\xspace}

\def\sutd{SUTD\xspace}

\def\SSS{S3\xspace}
\def\hamids{HAMIDS\xspace}
\def\hamidsLong{HierArchical Monitoring Intrusion Detection System\xspace}
\def\argus{ARGUS\xspace}

\def\SSSlong{SWaT Security Showdown\xspace}

\def\teamone{Team 1\xspace}
\def\teamtwo{Team 2\xspace}
\def\teamthree{Team 3\xspace}
\def\teamfour{Team 4\xspace}
\def\teamfive{Team 5\xspace}
\def\teamsix{Team 6\xspace}
\fi

\makeatletter
\def\@copyrightspace{\relax}
\makeatother

\begin{document}

% LICENSE {{{2
% \setcopyright{acmcopyright}
%\setcopyright{acmlicensed}
%\setcopyright{rightsretained}
% DOI
% \doi{10.475/123_4}
% ISBN
% \isbn{123-4567-24-567/08/06}
%Conference
% \conferenceinfo{ASIACCS '17}{TBD}
% \acmPrice{\$15.00}
%
% --- Author Metadata here ---
% \conferenceinfo{ASIACCS}{'17}
%\CopyrightYear{2007} % Allows default copyright year (20XX) to be over-ridden
%- IF NEED BE.
%\crdata{0-12345-67-8/90/01}  % Allows default copyright data
%(0-89791-88-6/97/05) to be over-ridden - IF NEED BE.
% --- End of Author Metadata ---

% TITLE {{{2
% \title{On a Cyber-Physical CTF}
% \title{Holistic CTF-Style Security Education and Research for Industrial
%Control Systems}
% \title{S3: Design and Results of CTF-Style Security Assessment of an ICS}
%MO: Not sure about using S3 for blinding purposes
%\title{Gamifying Education and Research on ICS Security: Design,
%Implementation and Results of \SSS}
\title{Gamifying Education and Research on ICS Security: Design,
Implementation and Results of \SSS}
% \subtitle{Anonymized for double-blind review}
%

\numberofauthors{5}
\date{1 November 2016}
\author{
    \alignauthor Daniele Antonioli\\
    \alignauthor Hamid Reza Ghaeini\\
    \alignauthor Sridhar Adepu\\
    \and 
    \alignauthor Mart\'in Ochoa\\
    \alignauthor Nils Ole Tippenhauer\\
    \and 
    \email{\{daniele\_antonioli, sridhar\_adepu, martin\_ochoa,
    nils\_tippenhauer\}@sutd.edu.sg}, \\
    \email{ghaeini@acm.org} \\
    \smallskip
    \sharedaffiliation
%    \affaddr{Information Systems Technology and Design (ISTD)}  \\
    \affaddr{Singapore University of Technology and Design}
}
% \author{}
% Daniele, Hamid, Sridhar, Martin, Nils
%
% Just remember to make sure that the TOTAL number of authors
% is the number that will appear on the first page PLUS the
% number that will appear in the \additionalauthors section.
\maketitle

% ABSTRACT {{{1

\begin{abstract}

%    Authors colors: \martin{Martin}, \daniele{Daniele}, \nils{Nils},
%    \hamid{Hamid}

    In this work, we consider challenges relating to security for
    Industrial Control Systems (ICS) in the context of ICS security
    education and research targeted both to
    academia and industry. We propose to address those
    challenges through \emph{gamified} attack training and countermeasure
    evaluation.

    % In recent years, ICS security has gained increasing attention in the
    % security research community. Unlike traditional ``office'' security, in ICS
    % security domain-specific devices and protocols are the subject of research.
    % Due to related cost, effort to set up, and potential health risks, only few
    % practical security-focused ICS testbeds appear to exist so far. As result,
    % researchers often have a vague understanding of real-world system
    % architectures, and cannot experimentally test attacks and defenses. In
    % fact, bringing the traditional ICS and security communities closer has been
    % pointed out by several researchers as a fundamental societal challenge to
    % improve the security of these systems.  In this
    % work, we propose a number of solutions that address those problems. In
    % particular, we propose ways to educate users about real-world platforms
    % using real or simulated devices, and gamify ICS security testing.

    % event is a contribution too
    We tested our proposed ICS security gamification idea in the context of
    the (to the best of our knowledge) first Capture-The-Flag (CTF) event targeted
    to ICS security called
    \emph{\SSSlong (\SSS)}.
    % \footnote{Name of event anonymized for double-blind  review.}.
    % that we present as an additional contribution.
    %\SSS involved international teams of people
    %from academia and industry.
    Six teams acted as attackers in a security competition leveraging
    an ICS testbed, with several academic defense systems attempting
    to detect the ongoing attacks.
    % phases: online and live: scoring might be a contribution too
    The event was conducted in two phases. The online phase (a
    jeo\-pardy-style CTF) served as a training session.  The live
    phase was structured as an attack-defense CTF\@. We acted as
    judges and we assigned points to the attacker teams according to a
    \emph{scoring system} that we developed internally based on
    multiple factors, including realistic attacker models.
	%We gave
    %prices to everyone from the attacker teams because we think that
    %rewards are an additional boost to increase the final learning
    %experience.
	We conclude the paper with an evaluation and
    discussion of the \SSS, including statistics derived from the data
    collected in each phase of \SSS.

    % We implemented our ideas as part of the (to the best of our knowledge) first
    % Capture-The-Flag style event on a productive Industrial Control
    % System. Our experiment involved six attacker and six defender teams from
    % academia and industry world-wide. Attackers were provided with the
    % documentation of the target systems, and the CTF was composed of two
    % phases: an online qualifier and a live-phase. During the online phase the
    % attackers were exposed to simulations and remote-access to a physical water
    % treatment plant. On the live-phase of the experiment, teams deployed a
    % wider range of attacks during two days, under surveillance by the defender
    % teams. In this work we report on insights and lessons learned from our
    % experiment.

\end{abstract}

% KEYWORDS {{{1
%\keywords{Capture-The-Flag; Industrial Control Systems; Gamification; Security;}

% INTRO [1 page] {{{1
\section{Introduction}

% ICS, ICS security trends and problem
Industrial Control System (ICS) security is a major challenge because it
requires specific knowledge about domain-specific devices, industrial
protocols and general knowledge about traditional IT security
threats~\cite{urbina16fieldbus, Shahzad2015, liu2011false}.
A common trend in ICS is the shift
towards commodity computing platforms and communication channels,
e.g., TCP/IP based communication using Ethernet instead of
RS-485~\cite{galloway2013introduction}. That shift if motivated by
increased functionality, together with cost savings. In addition, even
smaller devices are increasingly connected to networks (\eg as part
of the Industry 4.0 paradigm~\cite{sadeghi15iiot}), and use access to
the Internet to report data or obtain updates.

% ics vs ict security training
Recently, it has been widely argued that one of the fundamental issues in
securing ICS lies in the cultural differences between traditional IT security
and ICS engineering~\cite{schoenmakers2013contradicting,TNOICS}. Therefore,
education has been advocated as a means of bridging the gap between these
cultures~\cite{luiijf2015cyber,TNOICS}. However, recent surveys indicate that
although general IT security education efforts have risen in ICS, there is
still need for more targeted education combining both security and ICS
specific knowledge~\cite{sansICS}.

% ICS testbeds as a solution
ICS testbeds constitute a convenient environment to study ICS security, however
their deployment is rare because of many reasons, such as cost and
manpower~\cite{cardenasAminLinHuangHuangSastry,weerakkodyMoSinopoli}.
Researchers are usually not able to get access to such testbeds, and those
willing to do research on ICS security are facing many problems, such as lack of
understanding of a real ICS and the inability to test (new) attacks and
countermeasures in a realistic setup. Another common issue in ICS security
is resulting from the intrinsic inter-disciplinary nature of the subject,
namely it is difficult to bring together people from different expertise
domains, such as control theory, information security, and engineering.

% ics difficult to access
% In recent years, ICS security has gained increasing attention in the
% security research community. Unlike traditional ``office'' security, in
% ICS security domain-specific devices and protocols are often the
% subject of research~\cite{urbina16fieldbus, Shahzad2015, MORE}. Due to related
% cost, effort to set up, and potential health risks, only few practical
% security-focused ICS testbeds appear to exist so
% far~\cite{cardenasAminLinHuangHuangSastry,weerakkodyMoSinopoli}. As a
% result, researchers often have a vague understanding of real-world
% system architectures, which makes it difficult for them to design targeted
% attacks and defenses, since often they cannot experimentally test them or gain
% inspiration by interacting with realistic systems.

% proposal
In this work, we propose a number of solutions to those problems, and we
elaborate one of them based on gamification in the context of ICS
security \emph{education} and \emph{research} targeted both to
\emph{academia} and \emph{industry}. We strongly believe that gamification
combined with access to real and simulated ICS is a key driver for ICS
security progress at all levels (from beginner to expert) and among
different professional communities (both academia and industry).

% In this work, we propose a
% number of solutions that address those problems. In particular, we propose ways
% to educate users about real-world platforms using real or simulated devices,
% and gamify ICS security testing. This approach is similar in spirit to the
% recent work proposed by~\cite{ruef2016} in the field of Secure Software
% Construction.

% review
We implemented our ideas as part of the (to the best of our knowledge)
first Capture-The-Flag style event on a productive Industrial Control
System held at our institution in the summer of 2016. The experiment
leveraged a realistic water treatment testbed that is available for
security research~\cite{mathur16swat}, together with a
simulation environment for some of the proposed challenges.

Our experiment involved six attacker teams from academia and industry
worldwide. Attackers were provided with the documentation of the
target systems, and the CTF composed of two phases: an online
qualifier and a live-phase. During the online phase the attackers were
exposed to simulations and remote-access to a physical water treatment
plant. On the live-phase of the experiment, teams deployed a wider
range of attacks during two days, with two academic attack detection
systems in place. In this work we describe the setup in detail, and
comment on insights and lessons learned.

%\daniele{Bullet list to be reviewed according to the abstract} MO: Looks good to me
We summarize our contributions as follows:

\begin{itemize}[noitemsep,nolistsep]
    \item We identify several issues that currently hinder the adoption of
        security in industrial control systems, and likewise prevent security
            researchers from becoming familiar with ICS.
    \item We propose a set of solutions to mitigate those issues, with a
        particular focus on \emph{gamified} interactive Capture-The-Flag events
            using simulated and real infrastructure.
    \item We present the design and implementation of a two-part CTF event in
        detail, and analyze its results.
\end{itemize}

This work organized as follows: in Section~\ref{sec:background}, we provide
brief background on Industrial Control Systems (ICS), the \swatlong, and
Capture-The-Flag (CTF) events.
In Section~\ref{sec:gamifying}, we present our problem statement and solution
ideas. One of those proposed ideas was implemented by us in two phases, which
are presented in detail in Section~\ref{sec:online} and
Section~\ref{sec:live}. We present an  analysis of the two events together
with lessons learned in Section~\ref{sec:discussion}. Related work is
summarized in Section~\ref{sec:related}, and we conclude the paper in
Section~\ref{sec:conclusions}.

% BACKGROUND [1.5 page] {{{1

\section{Background}
\label{sec:background}

In this section, we will introduce the relevant information about
Industrial Control Systems (ICS), the \swatlong
a real water treatment testbed and Capture-The-Flag (CTF) events.

% ICS {{{2
\subsection{Industrial Control Systems (ICS)}
\label{sec:ics}

% ICS def
Industrial Control Systems (ICS) are complex autonomous systems involving
different types of interconnected devices and an underlying physical process.
ICS are deployed to monitor and control different types of industrial
processes, such as critical infrastructure (water distribution and treatment),
and transportation systems (planes and railways).

% Device desc
It is useful to categorize ICS devices according to their main role in an ICS.
\emph{Control} devices, such as Programmable Logic Controllers (PLC), and
Human-Machine Interface (HMI), provides monitoring and programmable control
capabilities to the ICS.
\emph{Network} devices, such as industrial switches and
firewalls, provide the foundations to build a (complex) ICS network that may
include different segments, protocols, and topologies. Typically, the ICS
control network is (virtually) separated from other networks such as DMZ, and
office networks. \emph{Physical-process} devices, such as sensors and
actuators, directly interface with the underlying physical process producing
analog and digital signals that will be eventually converted and processed
by other devices.
Furthermore, an ICS might be viewed as a combination of
OT (Operational Technology) and IT (Information Technology) devices,
indeed it requires different expertises to be managed.

% ICS sec
ICS security is a major challenge for many reasons. The complexity and
diversity of devices involved in an ICS increases the attacker
surface, namely an attacker can attack both the cyber-part and the
physical-part of an ICS\@. Additionally, modern ICS are embracing
standard Internet communication technologies, such as TCP/IP based
industrial protocol, resulting in ICS that can be controlled (and
attacked) from the Internet. Finally, the software of ICS devices may
contain vulnerabilities for several common reasons, such as:
un-patched or impossible to patch legacy code, the absence of standard
security certifications for ICS devices, and lack of resources to keep
the ICS updated. We note that ICS often run proprietary licensed
Operating Systems, firmware, and management software.

Arguably, threats to ICS focus on impact to the physical world,
instead of attacks on confidentiality or integrity of information. As
such, the damage by such attacks is expected to cause financial cost
due to destroyed property and decreased operational availability of
commercial systems. Famous examples of attacks on ICS are
Stuxnet~\cite{falliere2011stuxnet} and the attack on a wastewater
treatment facility in Maroochy~\cite{slay2007lessons}.

%\sridhar{May be we can mention one paragraph about previous recent attacks
%explicitly? Which might provide intuition to the first time reader}\nils{if
%you have references, please add them}

% SWAT {{{2
\subsection{\swatlong}
\label{sec:swat}

% SWAT intro and stages
In this work, we leverage the \swatlong (\swat) for experimental
work. \swat is a state-of-the-art water treatment testbed opened at
our institution in 2015 that comprises six stages or subsystems:

\begin{enumerate}[noitemsep,nolistsep]
    \item \textbf{Supply and Storage} pumps raw water from the source to
        the Raw water tank.
    \item \textbf{Pre-treatment} chemically treats raw water controlling
        electrical conductivity and pH.
    \item \textbf{Ultrafiltration (UF) and backwash} purifies water using
        ultrafiltration membranes, collects ultra-filtrated water in the
        Ultra-filtration tank, and periodically cleans the UF membranes.
    \item \textbf{De-Chlorination} chemically and/or physically (UV light)
        removes chlorine from ultra-filtrated water.
    \item \textbf{Reverse Osmosis (RO)} purifies water using RO process,
        separates the result into permeate (purified) and concentrate
        (dirty) water.
    \item \textbf{Permeate transfer and storage} store permeate water into the
        RO permeate tank.
\end{enumerate}

\begin{figure}[tb]
    \centering
    \includegraphics[width=\linewidth]{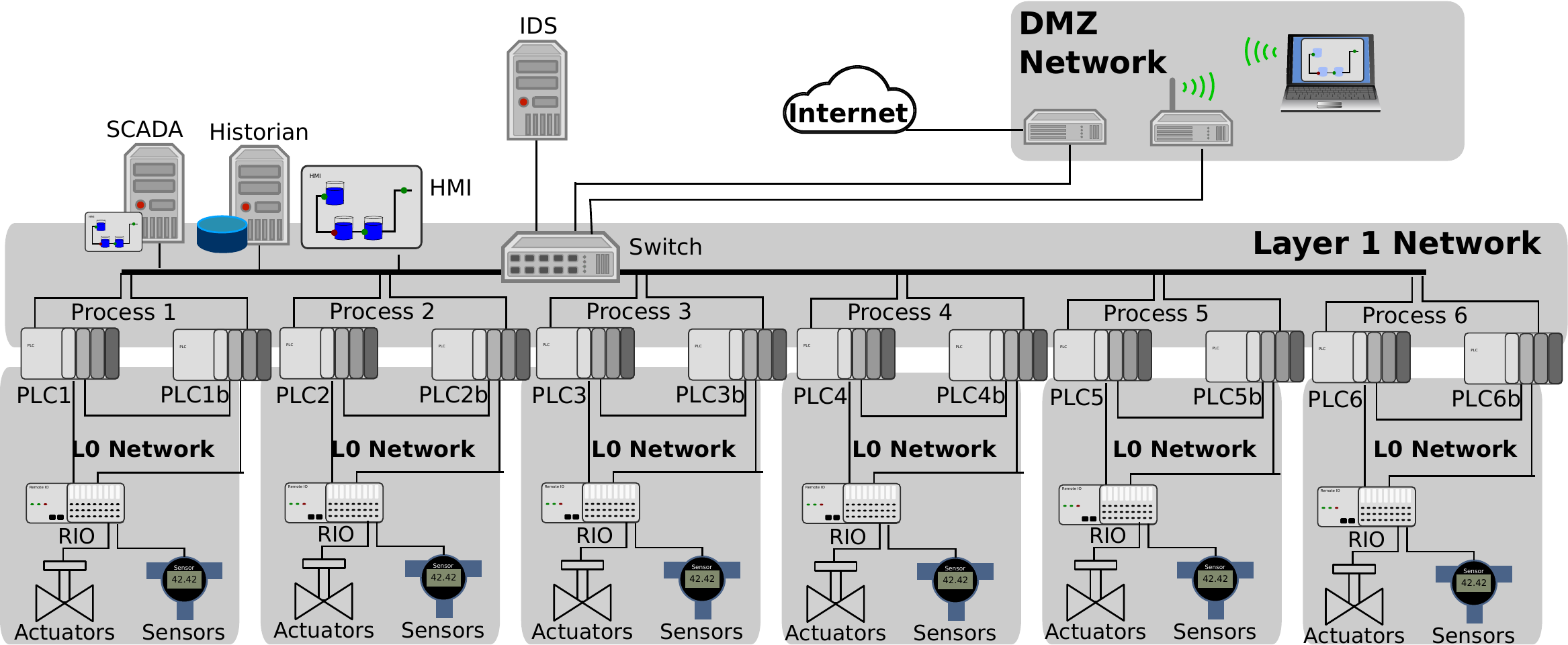}
    \caption{The \swat architecture.}
    \label{fig:swat}
\end{figure}

% SWAT fig (network) description
Figure~\ref{fig:swat} shows a schematic view of the \swat. Starting
from the bottom we can see six gray boxes representing the six water
treatment stages.  Each stage involves two PLCs, configured in
redundant mode, a Remote Input-Output (RIO) device, that translates
analog and digital signals, and a set of sensors and actuators. Each
stage's network is labeled as L0 (Layer 0), and it is an Ethernet
\emph{ring} topology built using the Device Level Ring (DLR)
protocol. Every PLC is connected to the L1 (Layer 1) network using a
conventional \emph{star} topology, a SCADA server, an HMI and a
Historian server. Other network segments access the \swat control
network through an industrial firewall, such as DMZ devices and
internet connected devices.  The spoken industrial protocol is \Enip,
that is an implementation of the Common Industrial Protocol
(CIP)~\cite{enip} based on the TCP/IP protocol stack.

% taken from Daniele's SWaT technical report
\swat is specified to process at least 5 US gallons/min of
permeate (purified) water, permeate water with pH from 6 to 7 and
electrical conductivity of at least 10 $\mu S / cm$, less than $50\%$
concentrated (dirty) water recirculation, and recover at least $70\%$
processed (total) water.

% CTF {{{2
\subsection{Capture-The-Flag (CTF) events}
\label{sec:ctf}

% ctf def
% Capture-The-Flag events are contests in the field of security where teams
% are challenged with finding and exploiting vulnerabilities in a wide range
% of categories. There exist two salient types of CTF: \emph{Jeopardy style},
% including categories such as reverse engineering, cryptography, web security,
% exploitation and forensics, and \emph{Red team/Blue team}, where teams compete
% to attack and defend an asset, such as a server.

% ctf def and types
Capture-The-Flag (CTF) are cyber-security contests organized worldwide by
Universities, private companies and non-profit organizations. CTF events can be
classified as: \emph{jeopardy-style} or \emph{attack-defense}. A
jeopardy-style CTF usually is hosted online and involves a set of tasks
to be solved divided by categories, such as cryptography and reverse
engineering. Each task is presented with a description, a number of hints and
an amount of reward points. The solution of a challenge involves finding (or
computing) a message (the flag) with a prescribed format, such as
\texttt{CTF\{my\_flag\}}, and submitting it to the CTF scoring system. An
attack-defense CTF, also called Red (attacker) team/Blue (defender) team,
is organized both offline and online where each team is given an identical
virtual machine containing some vulnerable services. The teams are connected
on the same LAN, and their goal is both to have a high service runtime and to
penetrate the services of the other teams, \eg finding and exploiting a
vulnerable service has two benefits: it allows a team to patch a service to be
more resilient against adversarial attacks, and to attack other teams
vulnerable service. Both jeopardy-style and attack-defense, CTF have time
constraints (realistic scenario) and the team who scores most points wins the
competition. The presented paper uses both online jeopardy-style and
live attack-defense CTF styles to augment the learning experience.

% ctf organization stats
Such events attract the
attention of both industrial and academic teams and currently enjoy increasing
popularity, as indicated by an established website in this community, listing
CTF competitions worldwide~\cite{ctftime}: this website lists 100 events being
held worldwide,
some of them with a long tradition such as the hacker-oriented DEFCON
CTF~\cite{defcon} and the academic-oriented iCTF~\cite{childers2010}.
Of the 10,000 teams listed in CTF time in 2016, some are
academic and others are composed of a heterogeneous mixture of security
enthusiasts, many of them security professionals.

% ctf as learning vector
CTF-like gamified security competitions are expected to help the ICS
security community in many
ways~\cite{eagle04capture,radcliffe07capture,vigna2003teaching}.  A
CTF is an hands-on learning experience and it can be used as an
educational tool, research tool, and as an assessment tool. Ideally,
both recruiters and candidates from academia and industry benefit
participating in CTF events as they exercise key
aspects of the ICS security domain such as knowledge of security
(recent) threats, teamwork, analytical thinking, development of (new) skills,
and working in a constrained environment. The gamification
aspect of a CTF allows the participant to express his or her full
potential, \eg attack/defend without fear of consequences or bad
marks.  CTF events have already been proposed as a means to enhance
security education and
awareness~\cite{eagle04capture,radcliffe07capture,vigna2003teaching}.
Although such events cover a wide range of security domains, to the
best of our knowledge they do not include so far the security of ICS.

% ctf design challenges

% DESIGN [2 page] {{{1
\section{Gamifying Education and \\Research on ICS Security}
\label{sec:gamifying}

We start this section by summarizing current challenge statements from
academia and industry, and we leverage them to set the problem statement of
this work. Then, we propose number of solution approaches. We focus on one of
them, and discuss how it could be implemented.

\subsection{ICS Current Security Challenges}
\label{sec:challenges}

In recent years, experts have argued extensively about the criticality of
securing Industrial Control Systems (ICS)s. Many have pointed out that one fundamental
challenge in achieving this task lies in cultural and educational differences
between the fields of (traditional) information security and ICS security.
% \Paragraph{Different Culture in Academia vs. Industry}
According to Schoenmakers~\cite{schoenmakers2013contradicting}:
\emph{``Differences in perspectives between
IT and OT specialists can cause security issues for control systems. It is
important for organizations to keep in mind that different values between
groups can influence the perception of issues and solutions.''}, which
emphasizes the cultural clashes still existing between traditional IT security
and ICS specialists.

Education and training have been advocated to bridge this gap, but there still
work to do in this domain.
Luijif~\cite{luiijf2015cyber} describes the security of ICS as a societal
challenge, and recommends:\emph{``Many of these challenges have to be
overcome by both end-users, system integrators and ICS manufacturers at the
long run: (\dots) proper education and workforce development''}.

% \Paragraph{CPS-Specific Security}
Despite the problem of education being widely acknowledged, according to a
recent report published by SANS Institute~\cite{sansICS}:
\emph{``It is clear from our results that most of our respondents hold security
certifications, but the largest number of these (52\%) is not specific to
control systems (\dots) IT security education is valuable, particularly with the
converging technology trends, but it does not translate directly to ICS
environments.''}

% \Paragraph{Potential for Academia and Industry collaborations}
In order to
effectively improve the security of ICS it is thus crucial to
educate researchers and practitioners such that they are able to understand the
subtleties and domain-specific requirements and constraints of security and
ICS\@. As recently pointed out by Luijif in~\cite{TNOICS}:
\emph{``(\dots) ICS and (office) IT have historically been managed by separate
organizational units. ICS people do not consider their ICS to be IT\@. ICS are
just monitoring and control functions integrated into the process being
operated. ICS people lack cyber security education. The IT department, on the
other hand, is unfamiliar with the peculiarities and limitations of ICS
technology. They do not regard the control of processes to have any
relationship with IT\@. Only a few people have the knowledge and experience to
bridge both domains and define an integrated security approach. Organizations
that have brought the personnel from these two diverse domains together have
successfully bridged the gap and improved the mutual understanding of both
their IT and ICS domains. Their security posture has risen considerably.''}

%In addition, we found that communication between the process engineering and
%computer science community can be difficult due to different terminology used
%to describe similar concepts~\cite{greep16achieving}.

% % PROBLEM STATEMENT {{{2
\subsection{Problem Statement}
\label{sec:problem}

With the challenges from Section~\ref{sec:challenges} as a high-level goal in mind,
in the following we discuss the problem statement and the proposed solutions.

% \Paragraph{Problem Summary}
Based on the literature and our experience, we think that traditional IT
(security) professionals need more information about the following topics:

\begin{itemize}[noitemsep,nolistsep]
    \item Common device classes, network topologies, and protocols used in
        ICS.
    \item Design methodologies best-practices and operational objectives in
        ICS.
    \item Physical processes specifications.
    \item Control theory models.
\end{itemize}

Furthermore, training for ICS (security) professionals can be
beneficial in the areas of:

\begin{itemize}[noitemsep,nolistsep]
    \item Common ``modern'' Internet communication
        technologies (Ethernet, IP, TCP, UDP, NAT).
    \item Common security challenges and standard
        solutions (MitM attacks, TLS, firmware and software update schedules).
    \item Standardization and specifications related to security
        products. % to link with countermeasure evaluation of S3
\end{itemize}

\textbf{Problem statement} How can we create an impactful educational
experience that addresses the aforementioned gaps?

% PROPOSED SOLUTION {{{2
\subsection{Proposed Solution Approaches}
\label{sec:solutions}

We now propose a number of approaches to alleviate the outlined problems. We
then present a solution that covers several of the proposed approaches.

\begin{enumerate}[noitemsep,nolistsep]
    \item A set of common use cases for ICS\@.
        In particular, the use cases would include a fictional or
        real physical process and details on the communication network
        topology.
    \item Interactive ICS testbeds, that allow
        users to familiarize themselves with the control devices and
        protocols used and interact with the underlying physical process.
    \item Practical training on ICS and information security.
    \item Testing of security solutions through external parties, and standard
        certifications.
\end{enumerate}
% \pagebreak

% S3 {{{2
\subsection{The \SSSlong}

In this work, we focus on the aspect of training and validation of applied
security skills for industry professionals and researchers. Gamification
in education has been advocated as a means to enrich the learning
experience~\cite{kapp2012gamification}. In particular, within IT security, the
implementation of CTF-like competitions have been argued to be advantageous for
education and training~\cite{vigna2003teaching}. Inspired by the gamified
nature of CTF, we propose the following approach.

Our goal is to create a realistic environment where participants are encouraged
to think out of the box. In real-life ICS settings, several intrusion detection
mechanisms are in place to safe-guard critical operations. A successful attacker
would have to bypass such systems in order to pose a threat, and simulating
such settings would stimulate a participant's ingenuity to attempt creative
attacks. On the other hand, if successful, such attacks will potentially
unveil limitations of the defense mechanism. Therefore,  we
propose to divide participants in a training event into two categories:
participants interested in developing defenses for ICS (defenders) and
participants interested in testing the security of ICS (attackers).

In order to get the most out of an interaction with a real ICS testbed, it is
important to learn fundamental concepts of ICS security. However, this learning
phase should be as hands-on and gamified as possible.  To this extent, we
propose an \emph{on-line} training phase, where attackers
get familiar with ICS concepts and a particular critical infrastructure by
means of a jeopardy-style CTF\@. Different from traditional CTFs, the challenges
are tailored to highlight ICS concepts and use realistic simulations of ICS
networks and remote interaction with ICS hardware. In this phase, attackers
also should be aware of the internal workings of common defense mechanisms in
place, and documentations there-of are shared with them.

After this preparation, in a \emph{live} phase attackers phase interaction with
a live system that is being monitored by defenders. In this setting, attackers
should have concrete goals to achieve (or \emph{flags} in CTF jargon), and
their scoring should be influenced by the number of defenses triggered during
their attack. In order to motivate attackers to perform more creative and
difficult attacks, different attacker models can be suggested to them (i.e.\
insiders with administration capabilities, outsiders with network access) and
the scoring can be adjusted according to the attacker model chosen.

Finally, participants will have access to statistics on their performance based
on a unified scoring system taking into account both phases. Attackers will
benefit from this experience since in order to solve the on-line and live
challenges they will have to go through several of the topics discussed in the
previous subsection. On the other hand, defenders will benefit by putting their
solutions to the test against creative attackers.
% more details on design, goals, evaluation. Ideally leadsup to \SSS of course

We have implemented the proposed concepts at our institution in 2016, under
the name \SSSlong. In the following two sections, we present the two
main phases of that
event, which represent the two target systems we introduced earlier: a)
online challenges using questions and simulated systems, and b) live events
using a real physical ICS\@. Due to organizational constraints, and in order to
maximize the learning experience, we decided to
limit the participants to \SSSlong to selected invited teams from academia
and industry, both for attacker and defender roles, for a total of 12 invited
teams (6 attackers, 6 defenders, of which 3 academic and 3 industrial teams
respectively). Teams were not limited in size, but only a maximum of 4 members
could participate physically in the live event whereas remaining team members
could join remotely.

% why restricted number of participants
%That is why we decided to limit the number of participating teams
%to twelve, with that setup we were able to control our (limited) amount of
%resources and to provide the best possible learning experience for the
%participants.

% ONLINE [2 page] {{{1
\section{online phase of \SSS}
\label{sec:online}

In this section we will present the \SSSlong online event, and the details
about its setup, and presented challenges. We will describe more in detail three set of
challenges from the MiniCPS, Trivia and Forensics categories. We conclude the
section with a summary of the collected results.

% SETUP AND CHALLENGES {{{2
\subsection{Online phase Setup and Challenges}

% intro: teams, schedule
The \SSSlong online phase involved six teams of attackers, three from
industry, and three from academia. We presented a total of
\emph{twenty} challenges in preparation for this phase, and we offered
to each team a limited time to access to \swatlong, and the required
documentation to get familiar with the \swat and \enip.  We organized
two sessions, each one 48 hours long, where three teams at a time
attempted to solve the challenges for a total amount of
$510$ points.

% challenges: novel domain
The \SSS online phase was structured as a \emph{jeopardy-style} CTF, and did
not require physical access to the \swat.  The main goal of this phase was to
provide an adequate training to the attacker teams (third goal from
Section~\ref{sec:solutions}).  Please refer to Section~\ref{sec:ctf} for more
information about Capture-The-Flag events.

Table~\ref{tab:online-challenges} summarizes the
proposed tasks. We presented twenty tasks divided into
five categories: MiniCPS, Trivia, Forensics, PLC, and Misc, for a total of
$510$ points. Each category exercised several ICS security domains, such as
Denial-of-Service, and Main-in-the-Middle attacks. It is important to notice
that categories such as MiniCPS, Trivia, and PLC are \emph{novel} in the domain
of traditional jeopardy-style CTFs. Following CTF design best-practices we
presented the challenges of each category in increasing order of difficulty
\eg: solving challenge $x$ helped to solve challenge $x+1$, and when
necessary, we gave hints \eg: you could use \texttt{toolx}
to accomplish a certain task. In general, we used the online event as a
training session to prepare the attacker teams for the \SSS live event that is
described in Section~\ref{sec:live}.

\begin{table}[htpb]
    \centering
    \caption{\SSSlong Online challenges summary: 20 tasks, worth 510 points.}
    \label{tab:online-challenges}
    \begin{tabulary}{0.45\textwidth}{rccL}
    \toprule
    \textbf{Category}  & \textbf{Tasks} & \textbf{Points} &\textbf{Security Domains}  \\
    \midrule
    MiniCPS   & 5 & 210 & network mapping, DoS, reconnaissance,
        MitM attacks in ENIP, tampering, tank overflow  \\
    \midrule
    Trivia    & 6 & 45 & \swat's physical process, devices and attacks  \\
    \midrule
    Forensics & 4 & 105 & packet inspection, processing and cryptography \\
    \midrule
    PLC       & 3 & 60  & ladder logic, code audit and development  \\
    \midrule
    Misc      & 2 & 90 & web authentication, steganography \\
    \bottomrule
    \end{tabulary}
\end{table}

% scoring system

\begin{figure}[tb]
    \centering
    \includegraphics[width=\linewidth]{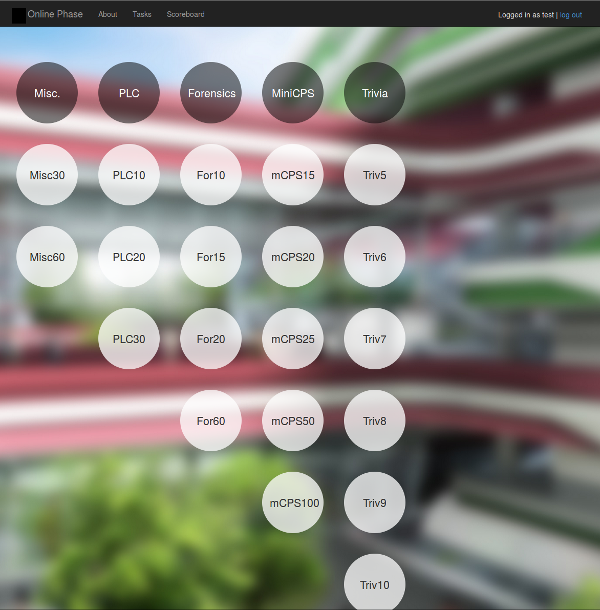}
    \caption{\SSS online challenges' web page.}
    \label{fig:ui}
\end{figure}

We built a Webapp to run the \SSS scoring system using the flask Python
framework~\cite{flask} (Figure~\ref{fig:ui} shows \SSS online challenges' web page).
The web pages were served over HTTPS, using Let's Encrypt~\cite{lets-enc}, and
a basic brute-force attempts detection mechanism based on user input logging
was put in place on the backend side.  A dedicated web page was showing a live
chart with the scores from all the teams. We offered live help with two
different channels: an IRC channel on \texttt{freenode.org}, and via email.
The following is an example of user interface interaction with our Webapp:
member of team A logs in to \SSS's Webapp (using the provided credentials),
she navigates to challenge X's Web page, then enters the flag on an HTML form.
If the flag is correct, she receives  $N$ reward points, otherwise a
submission error appears on screen.

In the following, we focus on the tasks related to MiniCPS, Trivia, and
Forensics categories, since they better illustrate the novel nature of the
challenges proposed, and then we will present a summary of the results from the
online event.

% MINICPS {{{2
\subsection{MiniCPS Category}

% why MiniCPS category, cite MiniCPS
The online phase presented five challenges in the MiniCPS category.
MiniCPS is a toolkit for Cyber-Physical System security
research~\cite{antonioli2015minicps}.  MiniCPS was used to
``realistically'' reproduce (simulate) part of the
\swatlong, including the hydraulic physical process, the
devices and the network. Each simulated instance accessed by the
attackers' teams was running on Amazon Web Services Elastic Compute
Cloud (AWS EC2), using an m3-type virtual machine (one instance per team).

\begin{figure}[b]
    \centering
    \includegraphics[width=\linewidth]{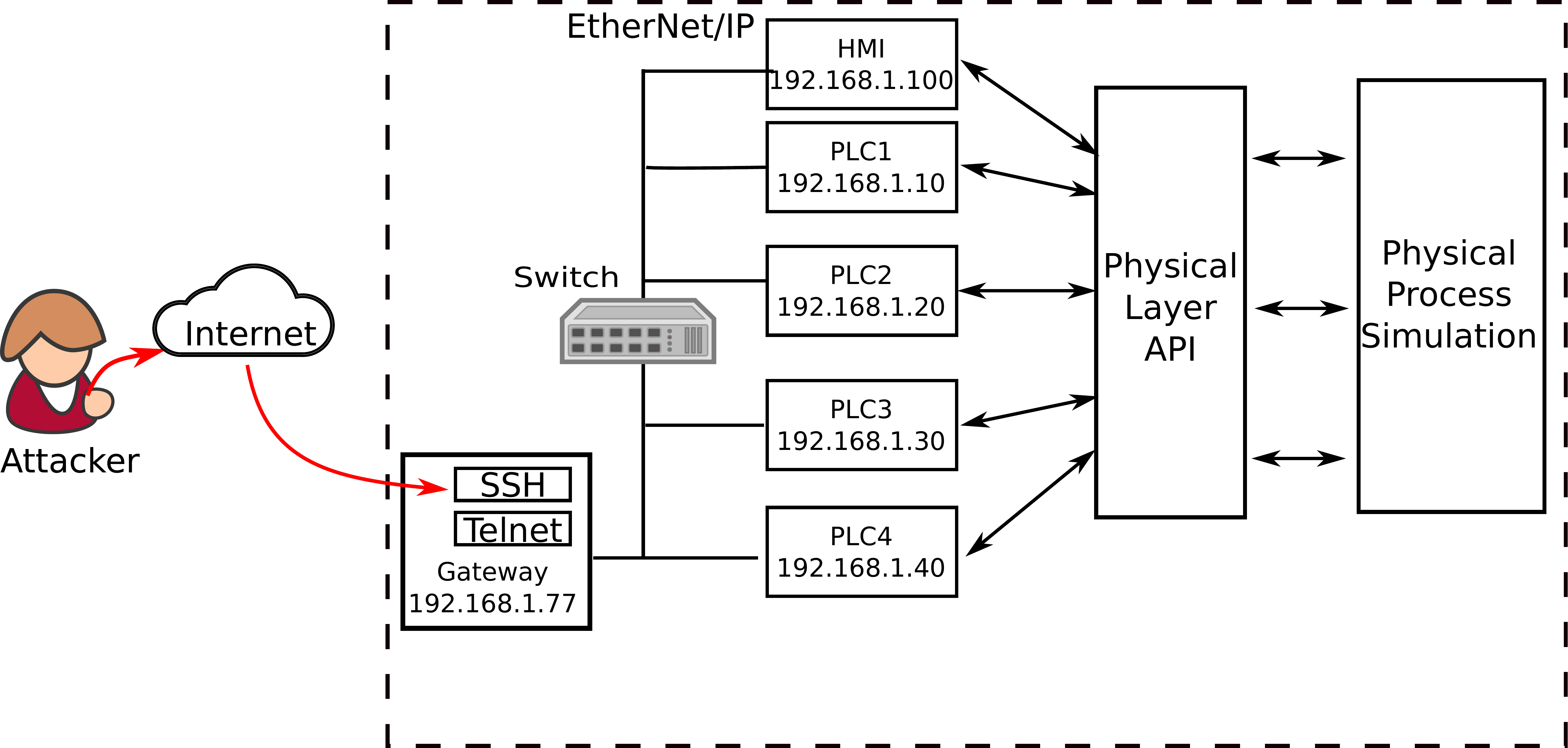}
    \caption{MiniCPS-based setup for online challenges.}
    \label{fig:minicps-honeypot}
\end{figure}

% explain figure, and tanks, cite honeypot
Figure~\ref{fig:minicps-honeypot} shows the simulation setup of a single
instance, that was replicated for all the six teams. Each attacking
team was provided with the credential to access an SSH server running on a
simulated chrooted gateway device. The attacker had access to the emulated
virtual control network that used the same topology, addresses (IP, MAC, net
masks), and industrial protocol (\Enip), of the \swat.

The attacker could interact with other simulated \swat's
devices in the star topology (four PLCs and an HMI), and alter the
state of the simulated water treatment process affecting the two simulated
water tanks (the Raw water tank and the Ultra-filtration tank).
For example, an attacker might send a
packet containing a false water level sensor reading of the Ultra-filtration tank
to the HMI, or a packet that tells to PLC2 to switch off the motorized valve,
that controls how much water goes into the Raw water tank.
As a side note, Figure~\ref{fig:minicps-honeypot}'s setup is part of an
internal project involving the development of novel honeypots for
ICS~\cite{antonioli2016honeypot}.

The following five paragraphs summarize each of the MiniCPS challenges, with
the attacker's goals and a reference solution:

% Net warm-up challenge
\Paragraph{Network warm up} The goal of the first challenge is to perform a passive
ARP-poisoning MitM attack between PLC2 and PLC3. The attacker has to perform a
network scanning to discover the hosts addresses and then use
\texttt{ettercap} to read the flag on the wire.

% ENIP warm-up challenge
\Paragraph{\Enip warm up} The goal of the second challenge is to read the flag
stored in PLC2's \enip server, and addressable with the name
\texttt{README:2}. The attacker has to understand which PLC owns the
\texttt{README:2} tag, and how to use \texttt{cpppo}, the suggested \enip's
Python library~\cite{cpppo}.

% Raw water tank overflow
\Paragraph{Overflow the Raw water tank} The goal of the third challenge is to
overflow the simulated Raw water tank. The attacker has to understand the
simulated dynamic of a water tank \eg who drives inflow and outflow, and
tamper with the
correct actuators to increase the water level above a fixed threshold. Some
hints were given to explain the binary encoding \eg use $m/n$ to switch ON/OFF
a water pump and OPEN/CLOSE a motorized valve.

% DoS HMI
\Paragraph{Denial of Service HMI} The goal of the fourth challenge is to
disrupt the communication (Denial-of-Service) between the HMI and PLC3, and
then change a keep-alive tag value to $3$ on PLC3 \enip's server. In normal
working condition the keep-alive tag is
periodically set by the HMI to $2$. Given the knowledge acquired from the
previous three challenges, the attacker has to perform an active MitM attack
that drops all the packets between the HMI and PLC3. Notice that,
it is not sufficient to just write the required keep-alive tag value on
PLC3 \enip's server.

% Ultrafiltration overflow
\Paragraph{Overflow the Ultra-filtration tank} The goal of the fifth challenge
is to overflow the Ultrafiltration water tank. The attacker has to reuse a
combination of the previously used techniques to set up an active MitM attack
using custom filtering rules \eg use \texttt{ettercap} and
\texttt{etterfilter}.

% TRIVIA {{{2
\subsection{Trivia Category}
The online phase presented six challenges in the Trivia category. The
Trivia challenges were intended for the attackers to understand the
plant structure, behavior, and the defense mechanisms. The knowledge
gained from these challenges is expected to be of use to the attackers
in other phases of the event.  In the remainder of this section, we
briefly describe the six challenges, their goals, and the steps needed
to capture the flags.

The trivia challenges can be divided into two types, the first type
involved the knowledge on \swat, and the second type involved research
papers on \swat.

\Paragraph{Knowledge on \swat} Three challenges fall under this
category. The goals of these challenges are to focus on the physical
process of \swat~\cite{swat}, the control strategy of \swat, and
the set points of the sensors and actuators. Following are the details
regarding the challenges.

{\bf Trivia 1} The goal of the first challenge is to identify the
analyzer that is used by the PLCs to control a specific dosing
pump. In order to identify the device, the participant has to
understand the control strategy of the particular dosing pump. As the
PLC uses a number of different inputs to control the dosing pump, the
participant has to trace the signals and identify the particular
analyzer.

{\bf Trivia 2} The goal of the second challenge is to find out the set
point that triggers the start of the backwash process. During the
filtration process, small particles clot the Ultrafiltration membrane. To
remove them and clean the Ultrafiltration membrane, a backwash process is
started after reaching a specific threshold. In order to answer this
challenge, the participant needs to revise and  understand the backwash
process.

{\bf Trivia 3} The goal of the third challenge is to identify the set point of
the hardness analyzer used by a PLC to shut down the RO filtration. The
hardness analyzer measures the water hardness in \swat. The set point is a
desired value of a particular sensor which is used by the PLC to control  the
plant. In the current scenario, when hardware analyzer exceeds desired value,
PLC shuts down the RO filtration.  In order to answer this challenge, attacker
should understand the set points and control strategy of the RO process.

\Paragraph{Research papers on \swat} The remaining three challenges
fall under this category.  The goals of these challenges are to raise
awareness about ICS attacks techniques and their classification in the
context of \swat. We selected the following three papers:
\cite{adepu2016investigation, kangAdepu, adepuMathurSGCRC2016} as
reference attack vectors targeting ICS\@. Following are
the details of those challenges.

{\bf Trivia 4} The goal of this challenge is to familiarise the attacker with
possible attacks on \swat and potential impact of those attacks on \swat. We
provided a research paper~\cite{adepu2016investigation} that presents an
experimental investigation of cyber attacks on an ICS\@. In order to answer
the challenge, the participant needed to read the paper and understand it.

{\bf Trivia 5} The goal of this challenge is to familiarise the
participant with a security analysis of a CPS\@. We provided another
research paper~\cite{kangAdepu} that presented a security analysis of
a CPS using a formal model. In order to answer the challenge attacker
should read the paper and understand it.

{\bf Trivia 6} The goal of this challenge is to familiarise the
attacker with multi-point attacks on ICS\@. We provided a third research
paper~\cite{adepuMathurSGCRC2016} that discussed multi-point
attacks. A multi-point attack leverages more than one entry point,
e.g, two or more communications links, to disturb the state of an
ICS\@. In order to answer the challenge, the participant needed to read
the paper and understand it.

% FORENSICS {{{2
\subsection{Forensics Category}

The forensics challenges focused on network capture files, in particular
\emph{pcap} files, that are easy to process using programs such as
\texttt{wireshark} and \texttt{tcpdump}. The participant had to
learn how to process and extract information from a file containing
pre-recorded network traffic from an ICS\@. The target industrial protocol was
\Enip. We now provide details on three of the four challenges.

% The aim of the forensics part is to learn how to handle a recorded ICS
% traffic. The flag format was s3flag{} and we ask some questions inside the
% recorded ICS traffic about the position of an attack, the payload of
% \enip packets and so on.

\Paragraph{Identify the ICS hosts} The goal of the first challenge is
to perform an analysis of the ICS hosts inside a captured ICS network
traffic.  To achieve that goal, the attacker should search for the
hosts inside the captured traffic, classify them based on their IP
addresses, identify whether a host is inside the ICS network or not
and enumerate them.

\Paragraph{Finding the poisoning host} The goal of this challenge is
to search for a host that has performed an ARP poisoning
Main-in-the-Middle attack, inside a captured network traffic. Then,
the attacker will identify the start point and end point of captured
ARP poisoning attack inside the captured network traffic. As an example, the
flag of this problem will be the start and end TCP sequence number in the form
of \texttt{ascflag\{A-B\}}, where the A is the starting TCP sequence number and
B is the ending TCP sequence number.

\Paragraph{Understanding the CIP protocol structure} The goal of this
challenge is to find a particular pattern inside the payload of CIP messages.
In particular, the attacker has to recognize that a CIP payload contains
encrypted data and then he has to decrypt it. So, the attacker can decrypt the
ciphertext by performing a XOR it with a key included in the payload or
performing a brute-force.

% ONLINE RESULTS {{{2
\subsection{Results from the Online phase}

Table~\ref{tab:online-results} presents the final scores of each team,
the number of captured flags, and an estimation of the time spent
playing, computed as the difference between the last and the first
flag submitted by a team. As we can observe from the table, two teams
were able to fully complete all tasks, with Team 6 being by far the
most efficient.  On average teams spent $25.67$ hours to solve the
challenges (53\% of the maximum of 48 straight hours), with a standard
deviation of $13.06$ hours.  The teams scored an average of $268.83$
points (52.7\% of the maximum of 510).  We believe that both the time
invested and the percentage of challenges solved shows a notable
investment in the game, and provides evidence on the engagement
generated by the gamification strategy. In addition, we note a
correlation between the number of hours invested, and the points
achieved. In fact, when the outlier (Team 6) is removed, there is a
$0.97$ Pearson correlation coefficient (PCC) between time spent and
points achieved.

% \begin{table}[htpb]
% \small
%     \centering
%     \caption{\SSS Online Results summary.}
%     \label{tab:online-results}
%     \begin{tabulary}{0.5\textwidth}{lccccccc}
%     \toprule
% & \multicolumn{5}{|c|}{Flags}&&\\
%     \textbf{Team}  & MiniCPS & Trivia & Forensics & PLC & Misc &Score&\textbf{Time}   \\
%     \midrule

%     % lancaster username id: 2
%     \teamone   &&&&    & $250$ & $13$ & $30$  \\
%     \midrule

%     % e&y id: 6
%     \teamtwo  &&&&      & $510$ & $20$ & $44$  \\
%     \midrule

%     % siemens id: 15
%     \teamthree   &&&&   & $86$ & $7$ & $27$  \\
%     \midrule

%     % applied risk id: 14
%     \teamfour    &&&&   & $161$ & $10$ & $28$  \\
%     \midrule

%     % UIUC id: 4
%     \teamfive    &&&&   & $66$ & $7$ & $21$  \\
%     \midrule

%     % nus id: 13
%     \teamsix     &&&&   & $510$ & $20$ & $4$  \\
%     \bottomrule

%     \end{tabulary}
% \end{table}

% task id 13-17: Minicps
% task id 0-5: Trivia
% task id 8-12: Forensic
% task id 18-20: PLC
% task id 6-7: Misc
\begin{table}[htpb]
    \centering
    \caption{\SSS Online Results summary. Category names: MCPS=MiniCPS, T=Trivia, F=Forensics, P=PLC, M=Misc}
    \label{tab:online-results}
    \begin{tabulary}{0.5\textwidth}{lrccccccc}
    \toprule
& \multicolumn{5}{|c|}{Flags per category}&&&\\
    \textbf{Team}  & MCPS & T & F & P & M &$\Sigma$ Flags&Score&\textbf{Time}   \\
    \midrule

    % lancaster username id: 2
    \teamone      & 2 & 6 & 4 & 0 & 1 & $13$ & $250$  & $30$h  \\
    \midrule

    % e&y id: 6
    \teamtwo      & 5 & 6 & 4 & 3 & 2  & $20$ & $510$  & $44$h  \\
    \midrule

    % siemens id: 15
    \teamthree    & 0 & 4 & 2 & 0 & 1 &   $7$& $86$  & $27$h  \\
    \midrule

    % applied risk id: 14
    \teamfour     & 4 & 4 & 2 & 0 & 0 & $10$& $161$  & $28$h  \\
    \midrule

    % UIUC id: 4
    \teamfive     & 0 & 4 & 2 & 0 & 1 &   $7$ & $66$ & $21$h  \\
    \midrule

    % nus id: 13
    \teamsix     & 5 & 6 & 4 & 3 & 2 & $20$ &$510$  & $4$h  \\
    \bottomrule

    \end{tabulary}
\end{table}

% \begin{table}[htpb]
%     \centering
%     \caption{\SSS Online Results summary.}
%     \label{tab:online-results}
%     \begin{tabulary}{0.5\textwidth}{lccc}
%     \toprule
%     \textbf{Team}  & \textbf{Score} & \textbf{Flags} &\textbf{Hours spent}  \\
%     \midrule

%     % lancaster username id: 2
%     \teamone       & $250$ & $13$ & $30$  \\
%     \midrule

%     % e&y id: 6
%     \teamtwo       & $510$ & $20$ & $44$  \\
%     \midrule

%     % siemens id: 15
%     \teamthree     & $86$ & $7$ & $27$  \\
%     \midrule

%     % applied risk id: 14
%     \teamfour      & $161$ & $10$ & $28$  \\
%     \midrule

%     % UIUC id: 4
%     \teamfive      & $66$ & $7$ & $21$  \\
%     \midrule

%     % nus id: 13
%     \teamsix       & $510$ & $20$ & $4$  \\
%     \bottomrule

%     \end{tabulary}
% \end{table}

% final remark
To conclude the online event, we believe that the gamification factor
played an important role. Gamification helped us
to combine different categories in an unified ICS security theme, to
motivate the attacker teams to do their best to get the maximum
points, and to implicitly train them for the upcoming live phase.

% LIVE [2 page] {{{1
\section{Live phase of \SSS}
\label{sec:live}

As discussed in Section~\ref{sec:gamifying}, the goal of the online phase
was to prepare teams for the live phase of \SSS.
In this section we will present the \SSSlong live event, and the details
about its setup, and scoring system. Afterwards, we will describe two of the
academic detection mechanisms, \argus and \hamids, that were used during the
event. We conclude the section providing a summary of the collected results.
% \pagebreak

% LIVE SETUP {{{2
\subsection{Live phase Setup}
The live phase of \SSS was held at \sutd over the course of 2 days in July
2016. All six attacker teams that participated in the online phase
were invited. Each team was assigned a three hour timeslot, in which
it would have free access to \swat to test and deploy a range of
attacks, taking advantage of the knowledge gained during the online
phase presented in Section~\ref{sec:online}.  In addition, teams were
able to visit the \swat testbed for one working day, to perform
passive inspection before the event to prepare themselves.

The main goal of the live phase was two-fold: firstly, the teams would be
able to learn more about an actual ICS and its security
(second and third goals from Section~\ref{sec:solutions}).
Secondly, we would be able to test a number of
(internally developed) detection systems that were deployed in \swat to
evaluate and compare their performances
(fourth goal from Section~\ref{sec:solutions}).

% SCORING AND ATTACKER PROFILES {{{2
\subsection{Scoring and Attacker Profiles}

We designed the scoring system for the live phase with the following goals:

\begin{itemize}[noitemsep,nolistsep]
    \item Incentivise more technically challenging attacks.
    \item De-incentivise re-use of same attack techniques.
    \item Provide challenges with different difficulty levels.
    \item Relate the attack techniques to realistic attacker models.
    \item Minimize damages to the participants and the actual system.
\end{itemize}

We now briefly summarize the scoring system we devised. In general,
points were only be awarded if the attack result could be undone by the
attacker (to minimize the risks of permanent damages).

Equation~\ref{eq:score} defines how to score an attack attempt:

\begin{equation} \label{eq:score}
    s = g * c * d * p
\end{equation}

With $s$ being the final score, $g$ a value representing the base
value of the goal, $c$ a control modifier to value the level of
control the attacker has, $d$ a detection modifier, and $p$ the
attacker profile modifier. Most modifiers were in the range $[1,2]$,
while the base value for the targets was in the range $[100,200]$.

% goal factor
We now describe in detail the four modifiers from Equation~\ref{eq:score}.
Goals could be chosen from two sets: \emph{physical process goals} and
\emph{sensor data goals}.

% NOTE: potentially turn this into tables
\Paragraph{Physical Process Goals $g$} Control over actuators, and
physical process (water treatment):

\begin{itemize}[noitemsep,nolistsep]
    \item 100 points: Motorised Valves (open/close/intermediate).
    \item 130 points: Water Pumps (on/off).
    \item 145 points: Pressure.
    \item 160 points: Tank fill level (true water amount, not sensor reading).
    \item 180 points: Chemical dosing.
\end{itemize}

\Paragraph{Sensor Data Goals $g$} Control over sensor readings at
different components:
\begin{itemize}[noitemsep,nolistsep]
    \item 100 points: Historian values.
    \item 130 points: HMI/SCADA values.
    \item 160 points: PLC values.
    \item 200 points: Remote I/O values.
\end{itemize}

% modifier factor
\Paragraph{Control modifiers $c$}
The control modifier determined how precise control the attacker had. As
guideline the modifier was $0.2$ if the attacker could randomly (value and
time) influences the process, up to $1.0$ if the attacker could precisely
influence the process or sensor value to a \emph{target value} chosen by the
judges.

% detection factor
\Paragraph{Detection modifiers $d$}  Not triggering a detection mechanism
while the attack is executed would increase the detection modifier, using the
following formula: $2-x/6$, with $x$ the number of triggered detection
mechanisms.

% attacker factor
\Paragraph{Attacker profile modifiers $p$}
For each attack attempt, the attacking team had to inform the judges
about the chosen attacker model before the attack is started. The overall
idea is that a weaker attacker profile would yield a higher
multiplier. The attacker profiles were based
on~\cite{rocchetto16models} and the higher is the modifier the weaker is the
attacker model. The \SSSlong live event used three attacker profiles: the
cybercriminal, the insider, and the strong attacker.

The \emph{cybercriminal} model had a factor of $2$.
The cybercriminal was assumed to have remote control over a machine in the ICS
network, and was able to use own or standard tools such as \texttt{nmap}, and
\texttt{ettercap}.
The cybercriminal did not have access to ICS specific tools, such as Studio
5000 (IDE to configure \swat's PLCs), or access to administrator accounts.

The \emph{insider} attacker had a factor of $1.5$. It represented a
disgruntled employee with physical access and good knowledge of the system,
but no prior attack experience, and only limited computer science skills. In
particular, the attacker was not allowed to use tools such as \texttt{nmap} or
\texttt{ettercap}, but had access to engineering tools (such as Studio 5000),
and administrator accounts.

The \emph{strong} attacker effectively combined both other attackers,
resulting in the strongest available attacker model, and
yielded a factor of $1$.

Attackers could earn points for one or more attacks. If more than one
attack was successfully performed, the highest scores from each goal
was aggregated as final score. For example, if an attack on pumps was
successful both using the strong attacker model, for a total score $s$
of 130 points, and the cybercriminal attack model, for a total score $s$
of 200 points, then only 200 points would be counted for that attack goal
(attack a pump).

% SRIDHAR'S DETECTION {{{2
\subsection{Detection mechanisms}

As discussed in Section\ref{sec:gamifying}, as part of the design of our
approach we included academic and commercial detection mechanisms as means to
incentivate the creativity of attackers: the less detection mechanisms
triggered the more points obtained, as discussed in the previous subsection.
Also, the experience was designed to serve as feedback to the designers of
detection mechanisms when confronted with various human attackers and a wider
range of attack possibilities. In the following we emphasise two academic
detection mechanisms implemented at \sutd.

\subsubsection{Distributed detection system}

The distributed attack detection method presented in
\cite{blindadepuMathurAsiaCCS2016, blindadepu-ieeecomp}  was implemented in Water
Treatment Testbed as one of the defense methods used in the \SSS event. The
method is based on physical invariants derived from the CPS design. A ``Process
invariant,'' or simply invariant, is a mathematical relationship among
``physical'' and/or ``chemical'' properties of the process controlled by the PLCs
in a CPS\@. Together at a given time instant, a suitable set of such properties
constitute the observable state of \swat.  For example, in a water treatment
plant, such a relationship includes the correlation between the level of water
in a tank and the  flow rate of incoming and outgoing water across this tank.
The properties are measured using sensors during the operation of the CPS and
captured by the PLCs at predetermined time instants. Two types of invariants
were considered: state dependent (SD) and state agnostic (SA). While both
types use states to define relationships that must hold, the SA invariants are
independent of any state based guard while SD invariants are. An SD invariant
is true when the CPS is in a given state; an SA invariant is always true.

The invariants serve as checkers of the system state. These are coded and the
code placed inside each PLC used in attack detection. Note that the checker
code is added to the control code that already exists in each PLC\@. The PLC
executes the code in a cyclic manner. In each cycle, data from the sensors is
obtained, control actions computed and applied when necessary, and the
invariants checked against the state variables or otherwise. Distributing the
attack detection code among various controllers adds to the scalability of the
proposed method. During \SSS, the implementation was located inside the
Programmable Logic Controllers (PLCs)\cite{blindadepu-ieeecomp}.

% HAMID'S DETECTION {{{2
\subsubsection{The \hamids framework}

The \hamids (\hamidsLong) framework~\cite{ghaeini16hamids} was
designed to detect network-based attacks on Industrial Control
Systems.  The framework leverages a set of distributed Intrusion
Detection System (IDS) nodes, located at different layers (segments)
of an ICS network. The role of those nodes is to extract detailed
information about a network segment, combine the information in a
central location, and post-process it for real-time security analysis
and attack detection. Each node uses the Bro Intrusion Detection
System (IDS)~\cite{paxson1999}.

\begin{figure}[tb]
    \centering
    \includegraphics[width=0.8\linewidth]{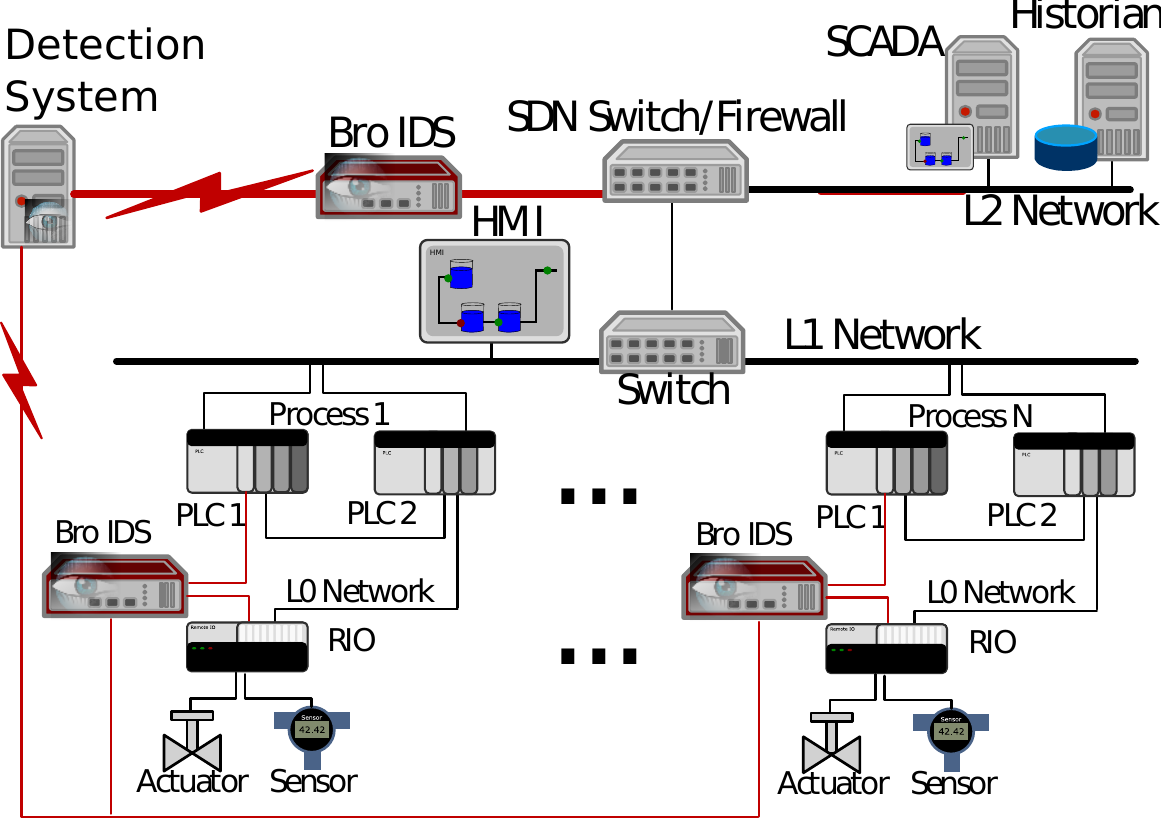}
    \caption{The \hamids framework instance: Bro IDS nodes are placed both at L0
    and L1 network segments of the \swat.}
    \label{fig:adaption}
\end{figure}

Figure~\ref{fig:adaption} shows our deployment of the \hamids
framework instance deployed in the \swat.
%Please refer to Section~\ref{sec:swat} for more information about \swat.
As we could see from Figure~\ref{fig:adaption}, each L0 (Layer 0) DLR
segment has an additional Bro IDS node that is collecting data flowing
from a PLC to the RIO device.  An additional Bro IDS node is connected
to a mirroring port of the L1 (Layer 1) industrial switch, and is
collecting the traffic in the L1 star topology.  Every Bro IDS node is
sending data to the central \hamids host, by means of a secure channel
(using SSH). Elasticsearch~\cite{elastic}, a distributed, RESTful
storage and search engine, is used to provide a scalable and reliable
information recording and processing.

The \hamids detection mechanism is entirely isolated from the ICS network, and
thus as part of the live event, the attackers were not able to have direct
access to the detection system. So the attackers will have hard times trying
to stop the detection mechanisms of the \hamids framework.  On the other side,
there are two ways to access data from a defender perspective: a Web interface
and a SQL API.

The web interface is a user-friendly interface that can be used by less
technical ICS operators and it is capable of listing all alarms generated by
the central node, and eventually help in detecting an ongoing attack.
The expert user can directly use the SQL API to query the central node, and
obtain more detailed data about the observed packets.

For the event, the framework was configured to present high-level
information about the status of the detection system, suitable for
non-expert defenders. Using the web user interface, the defender could
read the generated alarms related to triggered alarms due to observed
network traffic in the industrial control system. In addition, expert
defenders could read the detailed information about the ICS process by
using manual SQL queries to retrieve data for further analysis.

% LIVE RESULTS {{{2
\subsection{Results from the Live phase}

Table~\ref{tab:live-results} shows the final scores, the number of performed
attacks and the cumulative detection rate $d_{rate}$ of the live phase.
The cumulative detection rate was computed as the average
number of detection mechanisms triggered (considering only the two academic
detection mechanisms discussed before) in all successful
attacks by a given team.
% \martin{True?} \daniele{Yes} \sridhar{briefly presented
% in previous two sub sections}

%\daniele{do you like the idea of $d_{rate}$?} %Nils: yes

\begin{table}[htpb]
    \centering
    \caption{\SSSlong Live Results summary.}
    \label{tab:live-results}
    \begin{tabulary}{0.5\textwidth}{lccc}
    \toprule
    \textbf{Team}  & \textbf{Score} & \textbf{Successful Attacks} & \textbf{$d_{rate}$}  \\
    \midrule

    % lancaster username id: 2
    \teamone       & $666$ & $4$ & $1$ \\
    \midrule

    % e&y id: 6
    \teamtwo       & $458$ & $2$ & $1$ \\
    \midrule

    % siemens id: 15
    \teamthree     & $642$ & $3$ & $1$ \\
    \midrule

    % applied risk id: 14
    \teamfour      & $104$ & $1$ & $1$ \\
    \midrule

    % UIUC id: 4
    \teamfive      & $688$ & $5$ & $\frac{6}{5}$ \\
    \midrule

    % nus id: 13
    \teamsix       & $477$ & $3$ & $\frac{4}{3}$ \\
    \bottomrule

    \end{tabulary}
\end{table}

In order to show more in depth insights of the live phase, we now
provide details on several attacks that were conducted by the
participants during the \SSSlong live event (see
Table~\ref{tab:live-attacks}). We classify those attacks in two
types, the ``cyber'' attacks were conducted over the network using
either the cybercriminal, or the strong attacker model, while the
``physical'' attacks were conducted having direct access to the \swat
using either the insider, or the strong attacker model. We now
describe each of those attacks.

\begin{table}[htpb]
    \centering
    \caption{Live Attacks and Detections summary: \hspace{\textwidth}
    \Twdot \; = undetected, \Tdot \; = detected. }%NA = not applicable.}
    \label{tab:live-attacks}
    \begin{tabulary}{0.48\textwidth}{Lcccc}
    \toprule
    \textbf{Attack} & \textbf{Type} & \textbf{Score}   & \textbf{\argus} & \textbf{\hamids} \\
    \midrule

    SYN flooding (DoS) PLC                  & Cyber      & 396       & \Twdot  & \Tdot \\
    \midrule

    % L0 online manipulation          & 286       & \Twdot & \Twdot  \\
    % \midrule

    DoS Layer 1 network              & Cyber  & 104       & \Twdot & \Tdot  \\
    \midrule

    Tank level sensor tampering       & Physical       & 324       & \Tdot  & \Tdot  \\
    \midrule

    Chemical dosing pump manipulation & Physical  & 360       & \Tdot  & \Twdot \\
    \bottomrule

    \end{tabulary}
\end{table}

%\daniele{review attacks descriptions, and maybe remove one?}

% DoS SYN flooding: Siemens
\Paragraph{DoS by SYN flooding} The first attack was a cyber-attack, and the
attacker used the insider attacker model. The attacker had access to the
administrator account and associated tools. The attacker performed a SYN
flooding attack on PLC1's \enip server. SYN flooding is a Denial-of-Service
(DoS) attack, where the attacker (the client) continuously try to establish a
TCP connection, sending a SYN request to the \enip server, the \enip server
then responds with an ACK packet, however the attacker never completes the TCP
three-way-handshake and continues to send only SYN packets. As a result of
this DoS attack, the HMI's is unable to obtain current state values to
display, and would display 0 or * characters instead. Such effects would
impede the supervision of ICS in real applications.  However, the attack did
not interrupt or harm the physical process itself. The \hamids detectors was
able to detect the attack by observing the high number of SYN requests without
follow-up.  The \argus detector was not able to detect the attack, as the
physical process was not impacted.

% DDoS: Applied Risk
\Paragraph{DoS} The second attack was a cyber-attack, the attacker used
the cybercriminal attacker model. The attacker had access to the
network and attack tools. The attacker performed an ARP poisoning
Man-in-the-Middle attack, that redirected all traffic addressed to the
HMI. The redirected traffic was then dropped and prevented from being
received. The attack drove the HMI to an unusable state, and it took a
while to restore the system state after the attack. We did not allow
the attack to run long enough to affect the physical process. \hamids
detected the attack due to the changes in network traffic
(i.e. malicious ARP traffic, changed mapping between IP and MAC
addresses in IP traffic). In contrast, \argus did not detect the
attack, as the physical process continued to operate without impact.

% L0: Lancaster
% Nils: according to my notes, this was detected both by Hamids and by IC
% Daniele: detections taken from the score.xls, Lancaster page, first attack
\Paragraph{Tank level sensor tampering} The third highlighted attack involved an
on-site interaction with the system, the attacker used the strong attacker
model. The attacker focused on one of the L0 segments, and he
demonstrated control over the packets sent in the Ethernet ring. Indeed, the
attacker was able to alter the L0 traffic in real time, and manipulate the
communication between the PLC and the RIO\@.
\argus was able to detect the attack due to the sudden changes in reported
sensor values (bad data detector). In addition, the \hamids framework
detect the attack by observing the change in data reported from the PLC to the
SCADA (and potentially, in L0 as well).

% L0: EY
\Paragraph{Chemical dosing pump manipulation} The fourth attack was a
physical-attack, and the attacker used the insider attacker model. The
attacker was able to alter the chemical dosing in the second stage
(Pre-treatment) of the \swat by interacting directly with the HMI
interface, and overriding the commands sent by the PLC (that was set
in manual mode). The attack would have resulted in an eventual
degradation of the quality of the water, however we stopped the attack
before that case occurred.  \argus was able to detect the attack
because the updated setpoints diverged from their hard-coded counterpart in
the detection mechanism. The \hamids detection was
unable to detect this scenario as the network traffic did not show
unusual patterns or changes (as typical for attacks using the insider
model).

% DISCUSSION OF S3 [2 page] {{{1
\section{Discussion of \SSS}
\label{sec:discussion}

In the following section we present an analysis of a survey that we ran after
the \SSSlong, and some lessons learned during the online and live phases of
\SSS.

% \SSSlong is the product of TODO person-hours of work.

% COMPARISON {{{2
\subsection{Post-\SSS Survey Analysis}

% feedbacks
After the end of the \SSSlong event, we distributed an anonymous survey, both to
the attacker and defender teams. The survey asked targeted questions about
the online and the live phases, and for each question the team has to select a
score from 1 to 5; \eg 1 corresponded to full disagreement with the proposed
question, and 5 to full agreement. We present some insights from the survey
process, that overall resulted in a positive feedback from the participants and
in an interest to participate again in an security competition targeting ICS,
such as \SSS.

\begin{table}[htpb]
    \centering
    \caption{\SSSlong Attacker Survey summary: apart from *, scores from 1
    (fully disagree) to 5 (fully agree).}
    \label{tab:attackers-feedback}
    \begin{tabulary}{\linewidth}{Lcc}
    \toprule
        \textbf{Question}  & \textbf{Mean} & \textbf{Std Dev} \\
    \midrule
    Was the level of difficulty appropriate for the on-line phase?        & $3.67$ & $1.52$  \\
    \midrule
    Was the level of difficulty appropriate for the live phase?           & 4& $1$ \\
    \midrule
    Was the scoring appropriate for the on-line phase?                    & $3.67$ & $0.57$  \\
    \midrule
    Was the scoring appropriate for the live phase?                       & $4$ & $1$  \\
    \midrule
    Were the information shared beforehand and the on-line phase
    useful for preparing the live event?                                  & $3.67$ & $1.53$  \\
    \midrule
    *In your opinion, how much time would be required to prepare
    the attacks in the testbed?
        &$8.67$ h & $2.31$ h \\
    \midrule
    How do you rank your overall experience in the \SSS event?              & $4.67$ & $0.57$  \\
    \bottomrule

    \end{tabulary}
\end{table}

% table desc
Table~\ref{tab:attackers-feedback} shows the statistics of the answers
received from the attacker teams. As it can be seen
from the table, the overall satisfaction level is high, however there
are two low scores, one regarding the difficulty level of the online
event, and the other regarding the information shared beforehand. Most
of the teams commented that the time given to prepare the attacks in
the live phase was not enough.  Indeed, more time to interact with the
system is ideal and we will considering organizing future
events. However on the one hand, we think that strict time constraints
increase the level of realism of the competition, \eg an attacker who
penetrated a guarded ICS does not have unlimited time to perform his
attack.  On the other hand there are some inherent time and
organizational constraints: it is impossible to parallelize such a
competition, to avoid teams interfering with each other, and holding
such an event with even a restricted number of teams for several days
is resource-consuming (lab engineer and judges must be constantly
present).

In particular, we present two constructive feedbacks, one from
an attacker team, and one from a defender team, that we will take into account
in the next \SSSlong iteration:

\begin{itemize}[noitemsep,nolistsep]

    \item Attacker team: \emph{``There were some differences between the
        online and offline event, we made assumptions based on what we did
        online that ended up being wrong for the offline challenges. Overall
        it was a good lessons learnt for next time!''}

    \item Defender team: \emph{``I think that the attackers shouldn't have
        access to the HMI even with the
        insider profile. An insider should know the process and tags names and even
        have access to the HMI machine, but to perform an attack from the HMI itself
        is not only a boring attack that in the same manner the operator can do almost
        everything he wants, (\dots)
        % it is something that he can't do when someone else is
        % watching or if there are cameras in the control room, and it is not really a
        % cyber-attack.
        I think that the usage of the HMI should be only if the hackers
        manage to hack the HMI software and not from the user interface.''}

\end{itemize}

\medskip

Regarding the attacker comment, this is motivated by the fact that in the
simulations used in the live phase there were unavoidable simplifications of
the real system, which might have been misleading to some participants. In
future events we will better highlight this differences to better prepare teams.
The defender comment is interesting: indeed some defense products are aimed to
protect against malicious \emph{external} attackers, while trusting plant
operators with administrator privileges. However this highlights the
shortcomings of such mechanisms against malicious and powerful insiders.

% ONLINE LESSONS LEARNED {{{2
\subsection{Online phase: Lessons learned}

% \daniele{The following does not look like a lesson learned from the online
% event}
% By having a captured file of an ICS system, we could extract much useful
% information about the ICS process and ICS network structure. So, it is
% important to make the elements of ICS networks unreachable to an outsider
% adversary.

% MiniCPS lessons learned
We have learned a number of useful lessons from the MiniCPS related
challenges, and we will present three of them. Firstly, it is
hard to reproduce (simulate) the \swatlong: we were able to  provide partial
support of the \enip protocol, and replicate only the hydraulic part of the
\swat. Secondly, crash recovery and management is hard: during the competition
we suffered some downtime caused by attackers trying to use (legitimate)
brute-force attack techniques, and because of that some of them
wasted their time waiting for us to restore the system. Indeed, it is very
important to develop a simulation environment that is able to gracefully
shutdown, and restart automatically in most of the cases. Finally, side channel
attacks mitigation is hard. By side channel attack, we mean any attack that
uses a different channel from the ones intended by the competition.
Remember that, an attacker had to find one hole in our simulation environment
however we had to try to cover all of them. Indeed it is important to not
overlook the basic configuration of your system, even the parts not directly
related to the security competition. For example, remove unnecessary software,
and update all your software to the most recent (secure) version.

%Trivia
%\sridhar{Review please} \daniele{done}
We have learned mainly two useful lessons from the Trivia related challenges.
Firstly, it was helpful to present to the attacker general questions related
to the \swat's physical process, however presenting more advanced ones
would have helped the attacker in the live phase to prepare more
effective attacks. The same was true for the presented attack techniques. Given
that the majority of the participants was not from the ICS security domain, we
decided to present only basic attack techniques (already tested on the \swat),
however after the event we realized that we could have presented more
elaborated ones.

% The Trivia challenges intention is to make the attacker comfortable with the
% ICS domain, understand \swat, and different possible attacks on \swat. In the
% live event attackers expected to use the knowledge gained in these challenges.
% Firstly, we provided basic challenges in the Trivia, because most of the
% participants not from ICS domain.
% After the event we realized that it supposed to be advanced challenges so that
% attack teams get to know more of the \swat knowledge.
% Secondly, we provided
% basic challenges related to how attacks can be possible on \swat. The
% challenges are only providing basic level what kind of attacks possible. But
% these challenges missing how the advanced attacks possible on ICS.
% We believe, these lessons can be helpful for us to improve next event.

% LIVE LESSON LEARNED {{{2
\subsection{Live phase: Lessons learned}

% Live vs online
A key aspect of the live event is the interaction of the participants with
real devices, and with other people in a realistic environment.
Even though an online event is low-cost, scales better with the number of
participants, and presentes less risks in terms of safety, we learned that a
live event is an essential part of a security competition, and it was crucial
to include it in \SSS, even if it required more effort and risk management.

% Importance of background infotmation and prior access
Furthermore, we understood how important was to distribute documentation about
the \swat, and give access to it before \SSSlong.
During the live event it was easy to spot the teams who did not read the
provided documentation, and those teams got a lower score because they wasted
a lot of time to acquire the basic information about the testbed, such as:
number of devices, network topologies, and PLC programming.

% \sridhar{please review and edit...}
% Before the live event, we shared the \swat documents and organised online event.
% we have learned many lessons from live event of \SSS.  Firstly, which team
% explored more about the \swat, they are able to perform more in the live event
% than other teams. Which teams not explored much about the \swat are tried to
% get know during the event time, which reduces their time to prepare for an
% attack. It is explicitly shown that minimum knowledge \swat is important in
% order to perform intelligent attacks. Secondly, whichever teams have people
% from both the background automation and security, they are able to explore a
% bit quicker than other teams. Based on these lessons authors are suggesting
% future ICS CTF organisers in two aspects: Firstly, make sure that participants
% teams have enough domain knowledge in ICS and security. Secondly, as ICS CTF
% is new to the CTF community, by providing basic training and sufficient time
% before the event can make the event big success.

% cyber + physical
The major lesson that ICS professionals should learn from the live phase is
that ICS security consists of two intertwined parts: \emph{network} and
\emph{physical} security.
It is important to repeat that the attacker surface of an ICS is broad,
because it results in both cyber and physical attack.
Indeed, ICS security professionals should consider both the
cyber and the physical risk at the same time when dealing with ICS security.
% Without considering one part, the ICS system became vulnerable to many
% possible types of attacks,
% No sophisticated ICS network security elements
% shall not provide an acceptable level of ICS security without having ICS
% system safety in its mind.

% general
Finally, it is important to notice that designing a CTF is a hard
and time-consuming task.  Two key design aspects are the selection of
vulnerabilities and the scoring system. Let's use a jeopardy-style CTF
as an example. If the challenges are too hard then the newcomers may
lose interest after playing for a while, however, if the challenges
are too easy the participants will not enjoy the competition. Indeed
is crucial to design the challenges according to the audience
expertise, and to present novel threats rather than reusing material
from other passed events. The scoring system has to ``incentivize''
good behaviors and ``punish'' bad behaviors, according to the CTF
rules. However, it is impossible to predict, and stop attackers from
finding the novel technique to break the rules, and in some CTF (like
DEFCON) there are no rules at all!  In general, a good scoring system
has to be fair, automated and easy to understand, \eg participants
will have to focus on the CTF challenges, rather than try to find a
way to exploit an overcomplicated scoring system.

% RELATED WORK [0.5 page] {{{1
\section{Related work}
\label{sec:related}

In the following we review some
related efforts in gamification for security education.

% defcon CTF
There exists several popular CTF competitions, of which one of the most
established ones is DEFCON~\cite{cowan2003}. This is an annual hacking
conference organized by information security enthusiasts. The DEFCON CTF is
part of the main event, and it is  one of the most well known, and competitive
CTF events worldwide. DEFCON CTF has a jeopardy-style qualification phase,
and an attack-defense final phase, indeed its structure is simular to the one of
\SSSlong, however ICS security is not the main focus of DEFCON's CTF. Several
other similar CTF competitions are listed in~\cite{ctftime}.

In~\cite{vigna2003teaching} Vigna propose to use gamified live exercises
similar to CTFs to teach network security. The motiviations and philosophy
of this work are similar to ours, however the focus is on traditional IT and
network security (such as gaining root privileges in a webserver and steal data
from an SQL database).

% CTF alternatives: attack-only

Inspired by~\cite{vigna2003teaching}, in~\cite{childers2010} authors of the
iCTF event (organized by an academic instutition) presented two novel,
live, and large-scale security competitions. The first is called
``treasure hunt'' and it exercises  network mapping and multi-step network
attacks. The second is a ``Botnet-inspired'' competition and it involves
client-side web security, in particular Web browsers exploitation.
Unlike the presented paper, both competitions focus on traditional
client-server ICT network architectures and attack-only scenario.

% Web security (MIT/LL and Webseclab)
The MIT/LL CTF~\cite{werther11experiences}
was an attack-defense CTF with a focus on web application security. The main
goal of the event was to attract more people towards practical computer
security, lowering the barriers of a typical CTF that requires an extensive
background. The CTF takes inspiration from
Webseclab~\cite{bursztein2010}, a web security teaching Virtual Machine that
is packed with an interactive teaching web application, a sandboxed student
development environment, and a set of useful programs. Both are interesting
projects but they are not covering the ICS security domain, even though they
share some of the presented goals.

% bibifi
BIBIFI~\cite{ruef2016} is a cyber-security competition held mainly in academic
environments that adds secure development (Build-it) as a major component,
together with the attack (Break-it) and the patch (Fix-it) components.
This effort was targeted at improving secure software construction education,
and thus the exercises proposed in this competition do not cover the ICS
security domain.

% Offensive security
In~\cite{mink2010} Mink presents an empirical study that evaluates how
exercises based on gamification and offensive security approaches increase
both, the motivation and the final knowledge of the participants. Our work
tries to extend this message to ICS security, while the paper focuses on
traditional Information security.

In sum, to the best of our knowledge, we are the first to organize an academic
CTF style event on a realistic ICS testbed aimed at improving ICS security
training in academic and industrial environments.

%\nils{There is that other paper ``CTF not good/limited use for
%  education''. Can't find it anymore}

% CONCLUSIONS [0.5 page] {{{1
\section{Conclusions}
\label{sec:conclusions}

In this work, we discussed problems faced by security experts and ICS
engineers. In particular, security experts require easy platforms to
learn about ICS in general, and practise applied attacks and
defenses. In addition, ICS engineers often require additional training
in offensive and defensive security techniques. We proposed to use
gamified events such as online and live challenges to mitigate those
problems. We presented the design and implementation of two events
(online and live) that were conducted together by us in 2016,
leveraging a realistic ICS plant. To our best knowledge, this event
was the first attempt to gamify security education, using both live
and virtual ICS testbeds. Overall, the six participating teams
submitted 77 correct flags in the online phase of the \SSS event. In
the live phase, the participating teams performed 18 successful
attacks, of which most were detected by at least one of our detection
mechanisms. We provide a summary of challenges for both
events, and the achieved solutions by the participants, describe some
of the design choices made, e.g., relating to attacker profiles, goals
and scoring for the live event. The presented work should provide a
foundation to enable others to run similar events in the future.

% ACKN  {{{1
\section*{Acknowledgements}

We thank Kaung Myat Aung and all the staff from iTrust for their
support and contributions for event organization and
management. We thank all the attacker and defender teams for their
participation and valuable feedback.

% BIBLIOGRAPHY {{{1
\balance
\bibliographystyle{plain}
\bibliography{bibliography}

\end{document}